\documentclass[aps,pra,twocolumn,10pt]{revtex4-1}

\usepackage[utf8]{inputenc}
\usepackage{amsmath}
\usepackage{stmaryrd}
\usepackage{color}
\usepackage{mathrsfs}
\usepackage{yfonts}
\usepackage{rotating}
\usepackage{graphicx}
\usepackage{amsfonts}
\usepackage{mathrsfs}
\usepackage{amsmath}
\usepackage[american]{babel}
\usepackage{wasysym}
\usepackage{slashed}
\usepackage{braket}
\usepackage{graphicx}
\usepackage[caption=false]{subfig}
\usepackage{dsfont}
\usepackage{amsthm}
\usepackage{youngtab}

\usepackage{natbib}
\PassOptionsToPackage{breaklinks}{hyperref}
\usepackage[colorlinks,allcolors=black]{hyperref}

\theoremstyle{definition}

\usepackage{cleveref}
\crefname{proposition}{Proposition}{Propositions}
\crefname{theorem}{Theorem}{Theorems}
\crefname{definition}{Definition}{Definitions}
\crefname{lemma}{Lemma}{Lemmas}
\crefname{figure}{Figure}{Figures}
\crefname{corollary}{Corollary}{Corollary}
\crefname{conjecture}{Conjecture}{Conjectures}
\crefname{section}{Section}{Sections}
\crefname{appendix}{Appendix}{Appendixes}
\crefname{observation}{Observation}{Observation}
\crefname{remark}{Remark}{Remark}
\crefname{example}{Example}{Examples}
\crefname{equation}{Eq.}{Eqs.}
\crefname{table}{Table}{Tables}

\usepackage{tikz}

\usepackage{amssymb}
   \makeatletter
     \renewcommand\@make@capt@title[2]{%
      \@ifx@empty\float@link{\@firstofone}{\expandafter\href\expandafter{\float@link}}%
       {\textbf{#1}}\@caption@fignum@sep#2\quad}%
     \makeatother
\makeatletter
\renewcommand{\fnum@figure}{\textbf{Figure~\thefigure}}

\usepackage{mathtools}
\newcommand{\g}[0]{\gamma}
\newcommand{\al}[0]{\alpha}
\newcommand{\be}[0]{\beta}

\newcommand{\dd}[0]{\partial}
\newcommand{\la}[0]{\lambda}
\newcommand{\om}[0]{\omega}

\newcommand{\bs}[1]{\textbf{#1}}
\newcommand{\ea}[1]{\begin{align}#1\end{align}}
\newcommand{\eq}[1]{\begin{equation}#1\end{equation}}

\newcommand{\ma}[1]{\mathcal{#1}}

\usepackage{cancel}

\begin{document}

%%%%%%%%%%%%%%%%%%%%%%%%%%%%%%%%%%%%%%%%%%%%%%%%%%%%%%%%%%%%%%%%%%%%%%%%%%%
%%%%%%%%%%%%%%%%%%%%%    Title and Affitialitions     %%%%%%%%%%%%%%%%%%%%%
%%%%%%%%%%%%%%%%%%%%%%%%%%%%%%%%%%%%%%%%%%%%%%%%%%%%%%%%%%%%%%%%%%%%%%%%%%%

\title{Vacuum Energy from Qubit Entropy}
\author{Gon\c{c}alo M. Quinta}
\email{goncalo.quinta@tecnico.ulisboa.pt}
\affiliation{Physics of Information and Quantum Technologies Group, Instituto de Telecomunica\c{c}\~{o}es, Lisboa, Portugal}
\author{Antonino Flachi}
\email{flachi@phys-h.keio.ac.jp }
\affiliation{Department of Physics and Research \& Education Center for Natural Sciences, Keio University, 4-1-1 Hiyoshi, Yokohama, Kanagawa 223-8521, Japan}

%%%%%%%%%%%%%%%%%%%%%%%%%%%%%%%%%%%%%%%%%%%%%%%%%%%%%%%%%%%%%%%%%%%%%%%%%%%
%%%%%%%%%%%%%%%%%%%%%%%%%%%%%    Abstract     %%%%%%%%%%%%%%%%%%%%%%%%%%%%%
%%%%%%%%%%%%%%%%%%%%%%%%%%%%%%%%%%%%%%%%%%%%%%%%%%%%%%%%%%%%%%%%%%%%%%%%%%%

\begin{abstract}
We develop a non-conventional description of the vacuum energy in quantum field theory in terms of qubit entropy.
Precisely, we show that the vacuum energy of any non-interacting quantum field at zero temperature is proportional to the quantum entropy of the qubit degrees of freedom associated with virtual fluctuations. We prove this for fermions first and then extend the derivation to quanta of any spin. We also argue that essentially the same results are valid in the interacting case in the mean-field approximation and when the background is curved and static.
\end{abstract}

\maketitle

%%%%%%%%%%%%%%%%%%%%%%%%%%%%%%%%%%%%%%%%%%%%%%%%%%%%%%%%%%%%%%%%%%%%%%%%%%%
%%%%%%%%%%%%%%%%%%%%%%%%%%%    Introduction     %%%%%%%%%%%%%%%%%%%%%%%%%%%
%%%%%%%%%%%%%%%%%%%%%%%%%%%%%%%%%%%%%%%%%%%%%%%%%%%%%%%%%%%%%%%%%%%%%%%%%%%

\section{Introduction} The quantum vacuum is one of the building blocks underpinning the whole quantum field theory construct of Nature. Like its classical counterpart, the quantum vacuum does not contain physical particles, but differently from it, is not empty, rather filled with virtual ``zero-point'' fluctuations coming in and out of existence in accordance with the Heisenberg uncertainty principle. These virtual fluctuations are individually not observable, but their collective effect can have macroscopic implications, of which the Casimir effect in QED is, perhaps, one of the most spectacular manifestations \cite{Casimir:1948dh,Milonni:1994xx}.
%, although not the only \cite{Milonni:1994xx,Milonni:2000yj,Jaffe:2005xx}. 
A variety of experiments have conclusively observed macroscopic materializations of the quantum vacuum, at least in the incarnation of the Casimir effect down to the sub-micron range and with good accuracy (i.e., measured the quantum vacuum  force induced by the electromagnetic quantum fluctuations), providing a basis for an appreciation, at an operational level, of what %the energy of 
the quantum vacuum is \cite{Milton:2001xx,Bordag:2009xx,Lamoreaux:2004xx}. 

A more fundamental understanding %of the ``\textit{energy of the vacuum}'' 
remains elusive. The most evident example of this is illustrated by the cosmological constant problem \cite{Weinberg:1988cp}, i.e., the enormity of the contribution to the energy density of the Universe from the zero-point quantum field fluctuations, resulting in an energy density many orders of magnitude larger than what is observed. A closely related question has to do with the definition (and in practice also with the computation) of a \textit{finite} quantum vacuum energy for an interacting and %(perturbatively) 
possibly non-renormalizable quantum field theory, of which the canonically quantized formulation of gravity is the prime example. An intelligent scanning of the space of quantum field theories might be a way to better understand quantum vacuum effects, although in many cases the actual computation of vacuum energies is a difficult task. Taking an altogether different perspective may be a more rewarding path towards gaining new insights on phenomena associated with the vacuum. 

In this paper we propose an unconventional way to quantify the vacuum energy in quantum field theory in terms of quantum entropy of virtual fluctuations and illustrate a new connection underlying the information-theoretic structure of quantum field theory. More precisely, we make use of the qubit structure of spin-1/2 propagators (as $2 \times 2$ qubit systems described by a 4 dimensional matrix \cite{nielsen_chuang_2010}) to prove an identity relating the vacuum energy with the quantum entropy of those qubits. We further our arguments that the latter identity is valid for quantum fields of any spin, thereby hinting at some deeper fundamental connection between quantum fields and quantum information at the level of the quantum vacuum.

%%%%%%%%%%%%%%%%%%%%%%%%%%%%%%%%%%%%%%%%%%%%%%%%%%%%%%%%%%%%%%%%%%%%%%%%%%%
%%%%%%%%%%%%%%%%%%%%%%%    Vacuum energy of a fermionic field     %%%%%%%%%%%%%%%%%%%%%
%%%%%%%%%%%%%%%%%%%%%%%%%%%%%%%%%%%%%%%%%%%%%%%%%%%%%%%%%%%%%%%%%%%%%%%%%%%

\section{Vacuum energy of a fermionic field} As a starting point of our analysis, we shall consider fermionic quantum fields $\psi(x)$ of mass $m$, whose dynamics is dictated by the Dirac Lagrangian $\ma{L}=\bar{\psi}(x)(i \slashed{\dd} - m)\psi(x)$, where $x$ is the spacetime point of a Minkowsky spacetime and $\slashed{v}$ is the usual Feynman notation for $\slashed{v} = \g^{\mu} v_{\mu}$, with $\g^{\mu}$ being the Dirac matrices. The generalization to any spin will be addressed in turn. The partition function $Z$ for a fermionic field at temperature $T$ and zero chemical potential can be written as a path integral \cite{Kapusta:2006,Toms:2012xx}
\eq{\label{Zdef}
Z(T) = \prod_{n} \prod_{\bs{k}} \prod_{\al} \int i d\tilde{\psi}^{\dagger}_{\al;n}(\bs{k}) d\tilde{\psi}_{\al;n}(\bs{k}) \, e^{A[\tilde{\psi}]}
}
where
\eq{
A[\tilde{\psi}] = \sum_{n} \sum_{\bs{k}} i \tilde{\psi}^{\dagger}_{\al;n}(\bs{k}) (D_n(T))_{\al\nu} \tilde{\psi}_{\nu;n}(\bs{k})
}
and
\eq{\label{DnT}
D_n(T) = i \be \left(i\om_n+\g^0\g^j k_j +m\g^0\right)\,,
}
with $\al$ representing the spinor indices of the grassmanian variables $\tilde{\psi}$, $\be=1/T$, $\bs{k}$ is the spatial momentum and $n$ is an integer associated to the anti-periodicity in Euclidean time of the fermions yielding the Matsubara frequencies $\om_n = (2n+1)\pi T$. The partition function $Z$ can be calculated analytically \cite{Kapusta:2006,Toms:2012xx}, resulting in
\eq{\label{analyticZfermion}
\ln Z(T) = 2 \sum_{\bs{k}} \left[\be \om_{\bs{k}} +2\ln\left(1+e^{-\be\om_{\bs{k}}}\right) \right]\,,
}
where $\om_{\bs{k}} = \sqrt{|\bs{k}|^2+m^2}$. Recall that the average energy of the system can be obtained via the partition function using $\braket{E} = -\frac{\dd}{\dd \be} \ln Z(T)$ \cite{Kapusta:2006}, which results in
\eq{\label{E0def}
\braket{E(T)} = \sum_{\bs{k}} 2\omega_{\bs{k}}\left(\frac{e^{- \be \omega_{\bs{k}}}-1}{e^{- \be \omega_{\bs{k}}}+1}\right)\,.
}
The thermodynamic entropy $S(T)$, defined through $S(T) = \frac{\dd}{\dd T} (T\ln Z(T))$ \cite{Kapusta:2006}, gives
\eq{\label{Sdef}
S(T) = \sum_{\bs{k}} \frac{4\omega_{\bs{k}}\be}{e^{\be \omega_{\bs{k}}}+1} + 4\ln(1+e^{- \be \omega_{\bs{k}}})\,.
}
Combining Eqs.~(\ref{E0def}) and (\ref{Sdef}), it is possible to calculate the free energy $F$ as
\eq{\label{Fdef}
F(T) = \braket{E(T)}-T S(T)
}
or more simply, $F = - T \ln Z(T)$. In the zero temperature limit $T\to 0$, the thermodynamic entropy in Eq.~(\ref{Sdef}) goes to 0, and the free energy is all the energy available to the system, i.e. it becomes the vacuum energy $E_0$. More precisely, we have $E_0 \equiv \braket{E(0)} = F(0) = - 2\sum_{\bs{k}} \om_{\bs{k}}$, i.e. the usual expression as sum over the zero-point energies. This term is sometimes discarded by normal ordering, but in general (e.g., in the presence of boundaries or interactions) it cannot be naively eliminated, even in the non-relativistic limit \cite{Toms:2012xx,Edmonds:2023kon}. When boundary conditions are imposed, the momenta and, in turn, the frequencies are quantized, and the regularized summation gives, in the case of a renormalizable theory, a \textit{finite} Casimir or quantum vacuum energy. Irrespectively of the setup (e.g., of the boundary conditions), taking the limit of zero temperature $T \to 0$ of Eq.~(\ref{analyticZfermion}) leads to ${E_0 = - \lim_{T \to 0} T \ln Z(T)}$. On the other hand, Eq.~(\ref{Zdef}) can be used to derive the relation ${\ln Z(T) = \sum_{\bs{k}}\sum^{\infty}_{n=-\infty} \ln \det D_n(T)}$, which leads to
\eq{\label{E0det}
E_0 = - \lim_{T \to 0}\sum_{\bs{k}}\sum^{\infty}_{n=-\infty} T\ln \det D_n(T)\,.
}
The summations above are understood within an otherwise arbitrary regularization scheme. We also stress that finite temperature contributions do not add any additional divergence to the energy. The above formula will be our starting point to express the vacuum energy in terms of quantum entropy. Note that, despite the imaginary units present in the definition of $D_n(T)$, the explicit summations in Eq.~(\ref{E0det}) give a real value.

%%%%%%%%%%%%%%%%%%%%%%%%%%%%%%%%%%%%%%%%%%%%%%%%%%%%%%%%%%%%%%%%%%%%%%%%%%%
%%%%%%%%%%%%    Pseudo-qubit state of a virtual particle     %%%%%%%%%%%
%%%%%%%%%%%%%%%%%%%%%%%%%%%%%%%%%%%%%%%%%%%%%%%%%%%%%%%%%%%%%%%%%%%%%%%%%%%

\section{Pseudo-qubit state of a virtual fermion} The matrix $D_n(T)$ is very closely related to the fermion propagator $S_F(k)=(\slashed{k}+m)/(k^2-m^2)$, typically associated to the concept of virtual particles. We define the 4-momentum vector $k_n = (i\om_n,\bs{k})$, by virtue of the implicit Wick rotation present in the definition of the partition function $Z(T)$ in Eq.~(\ref{Zdef}). In particular, we have $k^0=i\omega_n$, i.e. the energy is quantized according to the Matsubara frequencies. One thus obtains $D_n(T) = i\be(k_n^2-m^2)S_F(k_n)\g^0$, or $D_n(T) = 4 i k^0 \be \check{\rho}(k_n)$ where we define the matrix
\eq{\label{rhoOriginal}
\check{\rho}(k) = \frac{(\slashed{k}+m)\g^0}{4k^0}.
}
The $4 \times 4$ matrix $\check{\rho}(k)$ has trace equal to unity and, for 4-momentum $k$ with real components, obeys $\check{\rho}^{\dagger}(k)=\check{\rho}(k)$. Its eigenvalues are $(1\pm r_{{\bs{k}}})/4$ (each with multiplicity 2), where we define the parameter
\eq{\label{rk}
r_{{\bs{k}}} = \frac{\om_{\bs{k}}}{k^0}\,,
}
thus $\check{\rho}(k)$ is semi-positive definite whenever $|r_{{\bs{k}}}| \leq 1$. The quantity $|r_{{\bs{k}}}|$ is the ratio between the on-shell and off-shell energies $\om_{\bs{k}}$ and $k^0$, respectively, so it basically defines the ``amount of virtuality'' of the fluctuations related to the fermionic propagator. Consequently, in the regime $|r_{{\bs{k}}}| \leq 1$ with $k$ real, in which case $\textrm{Tr}[\check{\rho}^2(k)]=(1+ r^2_{{\bs{k}}})/4<1$, one can associate $\check{\rho}(k)$ to a physically valid mixed density matrix in momentum space. This is expected since it would be contradictory to have a physically valid density matrix for a virtual particle in the entire phase space domain, a property which is characteristic of on-shell particles only. Outside of the range $|r_{{\bs{k}}}| \leq 1$, the matrix $\check{\rho}(k)$ loses its positive semi-definiteness and for complex momenta it will no longer be hermitian. As a consequence, we call $\check{\rho}(k)$ a \textit{pseudo-density matrix} of a virtual fermion as it generalizes the definition of density matrix to off-shell virtual quantum fluctuations related to the Dirac propagator.

An interesting fact is that $\check{\rho}(k)$ is a 2-qubit density matrix, where one of the qubits is related to spin projection and another one to energy sign \cite{Quinta:2023}. More precisely, one has that
\eq{
\check{\rho}(k) = \sum^{1}_{i,j,k,l=0} \rho_{ijkl}(k) \ket{i,j}\bra{k,l}
}
where the first qubit index in $\ket{i,j}$ selects positive or negative energy spinors and the second one selects spin up or down spinors. This can be intuitively understood from the fact that, in the computational basis, $\ket{00}=(1,0,0,0)^T, \dots, \ket{11}=(0,0,0,1)^T$, which correspond to spinors with distinct energy and spin projections in the rest frame. More precisely, from the energy and spin projectors
\ea{
\Lambda(\lambda_E) & = \frac{1+(-1)^{\lambda_E} \g_0}{2} \\
\Sigma(\lambda_s) & = \frac{1 -(-1)^{\lambda_s} \g_5 \g_3 \g_0}{2}
}
with $\lambda_E \in \{0,1\}$, $\lambda_s \in \{0,1\}$ and $\g_5 = i \g_0\g_1\g_2\g_3$ \cite{Bjorken:1964}, one can write the 2-qubit projection operator in the computational basis as
\eq{
\ket{\lambda_E \, \lambda_s}\bra{\lambda_E \, \lambda_s} = \Lambda(\lambda_E)\Sigma(\lambda_s)\,,
}
hence the association of $\lambda_E$ and $\lambda_s$ to qubits related to particle type and spin projection. The eigenstates of $\check{\rho}(k)$, which we denote as $\ket{\check\psi_{\la_E,\la_s}}$, are straightforwardly expressed in terms of the matrix
\eq{
W = \frac{(\om_{\bs{k}}+m)I-k_i \g^i \g_0}{\sqrt{2m(\om_{\bs{k}}+m)}}
}
which represents the Lorentz boost taking spinors from the rest frame to the frame with 4-momenta $k$. It is well known \cite{Bjorken:1964} that $W$ can be written in the exponential form
\eq{\label{expW}
W = e^{\al B}, \quad B = - \frac{k_i}{|\bs{k}|} \g^i \g^0\,,
}
where $\al$ is a hyperbolic angle defined via ${\cosh \al = \sqrt{\frac{\om_{\bs{k}}+m}{2m}}}$. In terms of Eq.~(\ref{expW}), the eigenstates $\ket{\check\psi_{\la_E,\la_s}}$ can be written as
\eq{\label{eigenRho}
\ket{\check\psi_{\la_E,\la_s}} = \sqrt{\frac{m}{\om_{\bs{k}}}} e^{\al B \g^0} \ket{\lambda_E \, \lambda_s}
}
with eigenvalues
\eq{
p_{\la_E,\la_s} = \frac{1+(-1)^{\la_E}r_{{\bs{k}}}}{4}\,.
}
The eigenvalues obey $\sum_{\la_E,\la_s} p_{\la_E,\la_s} = 1$ in general, although $p_{\la_E,\la_s}\leq 1$ only in the regime $|r_{{\bs{k}}}| \leq 1$. As expected, the eigenvalues only correspond to probabilities in the regime where $\check{\rho}(k)$ is well defined as a density matrix. One can also perform a Positive Partial Transpose test \cite{Horodecki:1996} on $\check{\rho}(k)$ to check its entanglement properties. Partial transposing it with respect to one of the qubits and calculating the respective eigenvalues, one finds the same eigenvalues $p_{\la_E,\la_s}$, which are larger or equal to 0 whenever $|r_{{\bs{k}}}| \leq 1$. One thus concludes that $\check{\rho}(k)$ is separable whenever it is well-defined as a density matrix.

The form of the eigenstates in Eq.~(\ref{eigenRho}), together with the fact that $\check{\rho}(k)$ is a mixed density matrix, suggests that $\check{\rho}(k)$ might have a thermal form. This is indeed the case, where it is possible to show that \cite{Quinta:2023}
\eq{\label{rhoT}
\check{\rho}(k) = \frac{e^{-\be_D H}}{\textrm{Tr}[e^{-\be_D H}]},
}
where
\eq{\label{Tvirtual}
\be_D = \frac{1}{2 k^0} \ln\left(\frac{1+r_{{\bs{k}}}}{1-r_{{\bs{k}}}}\right)
}
is a scalar quantity with inverse temperature units and ${H = -\left( m \g^0 - k_i \g^i \g^0 \right)/r_{{\bs{k}}}}$ is the associated Hamiltonian. Eqs.~(\ref{rhoOriginal})-(\ref{rk}) are valid for any general complex 4-momentum vector $k$. 

%%%%%%%%%%%%%%%%%%%%%%%%%%%%%%%%%%%%%%%%%%%%%%%%%%%%%%%%%%%%%%%%%%%%%%%%%%%
%%%%%%%%%%%%    Vacuum energy from quantum entropy     %%%%%%%%%%%
%%%%%%%%%%%%%%%%%%%%%%%%%%%%%%%%%%%%%%%%%%%%%%%%%%%%%%%%%%%%%%%%%%%%%%%%%%%

\section{Vacuum energy from qubit entropy} Equations (\ref{E0det}) and (\ref{rhoT}) converge into a relation between the quantum vacuum energy and the qubit entropy of the virtual quanta. Indeed, the matrix $\check{\rho}(k)$ is both intimately connected to the fermionic propagator and has a natural description in terms of qubits, so it serves as a bridge between quantum fields and quantum information. In fact, it is possible to show that the vacuum energy in Eq.~(\ref{E0det}) can be written in terms of $\check{\rho}(k)$ as (c.f. the Supplemental Materials for a proof)
\eq{\label{lndetrho}
E_0 = - \lim_{T \to 0}\sum_{\bs{k}}\sum^{\infty}_{n=-\infty} T\ln \det \check{\rho}(k)\,.
}
In other words, the vacuum energy has a direct connection with the qubits of virtual particles. Given the pillar importance of the concept of entropy in Quantum Information and Thermodynamics, it is therefore strongly expected that some connection between vacuum energy and qubit entropy exists. Indeed, defining the \textit{von Neumann} entropy functional
\eq{
S[M] = -\textrm{Tr}[M \ln M]\,,
}
where $M$ is a general matrix, and using the relation ${\ln\left(\textrm{Tr}[e^{-\be_D H}]\right) = S[\check{\rho}(k)] + \be_D \om_{\bs{k}}}$, one obtains (c.f. Eq.~(\ref{finiteTresult}) of the Supplemental Material for a rigorous proof)
\eq{\label{SlnZ}
\sum_{\bs{k}} \sum^{\infty}_{n=-\infty} S[\check{\rho}(k_n)] = -\frac{\ln Z(T)}{4}
}
where the summations are understood within some regularization scheme. From Eqs.~(\ref{E0def})-(\ref{Fdef}) and Eq.~(\ref{SlnZ}), one obtains
\eq{
F(T)=T \check{S}(T)
}
and
\eq{\label{1stlaw}
\braket{E(T)} = T(S(T)+\check{S}(T))
}
where we define
\eq{\label{totalS}
\check{S}(T) = 4\sum_{\bs{k}} \sum^{\infty}_{n=-\infty} S[\check{\rho}(k_n)]\,.
}
The quantity $\check{S}$ can be interpreted as the total entropy originated from the qubit degrees of freedom of virtual particles. The overall factor of $+4$ is associated to the spin-$1/2$ degrees of freedom in $3+1$ dimensions (the $+$ sign originates from the anticommutation properties of the fields, one factor of 2 comes from summation over spin indexes and another factor of 2 comes from summing over particles and anti-particles). For bosons, the overall sign would become negative and the multiplicative factor would change according to the dimension of the scalar multiplet in question \cite{Kapusta:2006,Toms:2012xx}. In addition, Eqs.~(\ref{analyticZfermion}) and (\ref{SlnZ}) would have their signs reversed as well as the sign multiplying by the exponential in Eq.~(\ref{analyticZfermion}). Eq.~(\ref{1stlaw}) can be interpreted as the internal energy of the fermionic Fock space at finite temperature, with two entropic contributions: one related to thermodynamics and another one originated from the entropy of qubits degrees of freedom associated to virtual particles.

Taking the limit $T \to 0$ in Eq.~(\ref{1stlaw}), one obtains the identity for the vacuum energy 
\eq{\label{E0S}
E_0 = \lim_{T \to 0} T\check{S}\,,
}
which succinctly expresses the fact that the vacuum energy at zero temperature is the sum over phase-space of energetic contributions related to the entropy of qubit states $\check{\rho}(k_n)$ of all possible spin-$1/2$ virtual particles. 

Finally, although the summations in $n$ and $\bs{k}$ can be regularized in Eq.~(\ref{totalS}), the expression is infinite in the limit $T \to 0$. This is consistent with Eq.~(\ref{E0S}), where the combination $T\check{S}$ gives a finite result. For finite temperature, Eq.~(\ref{totalS}) is not divergent, provided regularization schemes are applied to make sense of the summations.

%%%%%%%%%%%%%%%%%%%%%%%%%%%%%%%%%%%%%%%%%%%%%%%%%%%%%%%%%%%%%%%%%%%%%%%%%%%
%%%%%%%%%%%%    Vacuum energy from quantum entropy     %%%%%%%%%%%
%%%%%%%%%%%%%%%%%%%%%%%%%%%%%%%%%%%%%%%%%%%%%%%%%%%%%%%%%%%%%%%%%%%%%%%%%%%

\section{Generalizations} 
%to any spin} 
The results of the previous section can be %Eq.~(\ref{E0S}) can be 
generalized to non-interacting quantum fields of any spin, using the higher spin equivalent of Eq.~(\ref{analyticZfermion}) for a general quantum field $\chi$ with zero chemical potential
\cite{Kapusta:2006}
\eq{
\ln Z_{\chi} (T) = \frac{N_{\chi}}{2} \sum_{\bs{k}} \left[\be \om_{\bs{k}} +2\ln\left(1+e^{-\be\om_{\bs{k}}}\right) \right]\,,
}
where $N_{\chi}$ is a real constant proportional to the degrees of freedom of $\chi$ ($N_{\chi}=4$ for spin $1/2$). From Eq.~(\ref{E0def}), we also know that ${E_0=-(N_{\chi}/2) \sum_{\bs{k}} \om_{k}}$. Using directly the result (c.f. Eq.~(\ref{T0limit}) the Supplemental Material)
\eq{\label{Setom}
\lim_{T \to 0} \sum^{\infty}_{n=-\infty} T S[\check{\rho}(k_n)] = -\frac{\om_{\bs{k}}}{2}
}
%(where a regularization scheme is again assumed) 
follows that Eqs.~(\ref{1stlaw}) and (\ref{E0S}) still hold, provided that the entropy in Eq.~(\ref{totalS}) is generalized to include the extra degeneracies derived from the spin degrees of freedom, i.e. that the entropy from virtual qubits is generalized to 
\eq{\label{finalS}
\check{S}(T) = N_{\chi} \sum_{\bs{k}} \sum^{\infty}_{n=-\infty} S[\check{\rho}(k_n)]\,.
}

A further extension concerns the case of interacting quantum fields propagating on a curved background. We consider for simplicity the case of interacting fermions in the chiral limit (i.e., with vanishing bare mass and an interaction of the form $G \left(\bar\psi\psi\right)^2$ where $G$ is the coupling constant) and on a curved, homogeneous maximally symmetric spacetime. At mean field level, fermions acquire an effective mass proportional to the chiral condensate, $M_{\mathrm{eff}} = G \langle \bar\psi \psi \rangle$ and the one-loop thermodynamic potential takes the form $Z=-\nu \ln \det \left(i \slashed{\nabla} - M_{\mathrm{eff}}\right) = - {\nu \over 2} \ln \det \left(\Box + M_{\mathrm{eff}}^2 +{R\over 4}\right)$, 
%
%\begin{eqnarray}
%Z&=&-\nu \ln \det \left(i \slashed{\nabla} - M_{\mathrm{eff}}\right) = - {\nu \over 2} \ln \det \left(\Box + M_{\mathrm{eff}}^2 +{R\over 4}\right)
%\nonumber
%\label{OmegaR}
%\end{eqnarray}
where the second equality originates from the iteration of the Dirac operator, applying the Schr\"odinger-Lichnerowicz-Peres formula \cite{Parker:2009uva,Gibbons:2019coc}; the curved space D'Alembertian $\Box$ acts on spinors and $\nu$ is a constant proportional to the number of  degrees of freedom \cite{Flachi:2014jra}.  Thus the effect of interactions (in the mean field approximation) and of curvature can be assimilated by a shift in the effective mass. This shift maintains the main results of Eqs.~(\ref{1stlaw}) and (\ref{finalS}) unaltered. The shift in the effective mass due to the curvature translates into a shift in $r_{{\bs{k}}}$, thus changing the region in which virtual fluctuations behave as physical real fluctuation (i.e., where the density matrix is hermitian, has unit trace and is semi-positive definite), becoming larger as a positive curvature increases. The opposite is expected to happen when the background is negatively curved.

%%%%%%%%%%%%%%%%%%%%%%%%%%%%%%%%%%%%%%%%%%%%%%%%%%%%%%%%%%%%%%%%%%%%%%%%%%%
%%%%%%%%%%%%%%%%%%%%%%%%%%%%    Conclusions     %%%%%%%%%%%%%%%%%%%%%%%%%%%
%%%%%%%%%%%%%%%%%%%%%%%%%%%%%%%%%%%%%%%%%%%%%%%%%%%%%%%%%%%%%%%%%%%%%%%%%%%

\section{Conclusions} We have presented a novel interpretation of the vacuum energy %of a non-interacting 
in quantum field theory in terms of the quantum entropy of qubits associated to virtual particles. We have explicitly shown this by taking fermions as starting point and observed that the same relation holds true for any spin. We have argued that both curvature and interactions do not change the essence of these results, at least in the mean-field approximation. In deriving these results, a crucial step was the definition of a pseudo-density matrix $\check{\rho}(k)$, directly linked to the propagator of virtual fluctuations. %In fact, the object $\check{\rho}(k)$ plays a fundamental role in establishing a connection between the vacuum energy and the quantum entropy of qubits related to virtual spin 1/2 degrees of freedom.

It is counter-intuitive that quantum information not only exists in the vacuum, but can also be interpreted as the source of its zero-point energy. This connection may have various implications and, in particular, it may suggest a quantum vacuum origin of entropic forces.

%%%%%%%%%%%%%%%%%%%%%%%%%%%%%%%%%%%%%%%%%%%%%%%%%%%%%%%%%%%%%%%%%%%%%%%%%%%
%%%%%%%%%%%%%%%%%%%%%%%%%    Acknowledgements     %%%%%%%%%%%%%%%%%%%%%%%%%
%%%%%%%%%%%%%%%%%%%%%%%%%%%%%%%%%%%%%%%%%%%%%%%%%%%%%%%%%%%%%%%%%%%%%%%%%%%

\section{Acknowledgements} G.Q. thanks the support from Funda\c{c}\~{a}o para a Ci\^{e}ncia e a Tecnologia (Portugal), namely through project CEECIND/02474/2018 and project EXPL/FIS-PAR/1604/2021 QEntHEP - Quantum Entanglement in High Energy Physics. A.F. acknowledges the support of the Japanese Society for the Promotion of Science Grant-in-Aid for Scientific Research (KAKENHI, Grant No. 21K03540).

%%%%%%%%%%%%%%%%%%%%%%%%%%%%%%%%%%%%%%%%%%%%%%%%%%%%%%%%%%%%%%%%%%%%%%%%%%%
%%%%%%%%%%%%%%%%%%%%%%%%%%%    Bibliography     %%%%%%%%%%%%%%%%%%%%%%%%%%%
%%%%%%%%%%%%%%%%%%%%%%%%%%%%%%%%%%%%%%%%%%%%%%%%%%%%%%%%%%%%%%%%%%%%%%%%%%%

%%%%%%%%%%%%%%%%%%%%%%%%%%%%%%%%%%%%%%%%%%%%%%%%%%%%%%%%%%%%%%%%%%%%%%%%%%%
%%%%%%%%%%%%%%%%%%%%%%%    Supplemental material     %%%%%%%%%%%%%%%%%%%%%%
%%%%%%%%%%%%%%%%%%%%%%%%%%%%%%%%%%%%%%%%%%%%%%%%%%%%%%%%%%%%%%%%%%%%%%%%%%%
\clearpage
\pagebreak
\newpage

\widetext
\begin{center}
\textbf{\large Supplemental Material for ``Vacuum Energy from Qubit Entropy''}
\end{center}
%%%%%%%%%% Merge with supplemental material %%%%%%%%%%
%%%%%%%%%% Prefix a "S" to all equations, figures, tables and reset the counter %%%%%%%%%%
\setcounter{equation}{0}
\setcounter{figure}{0}
\setcounter{table}{0}
\setcounter{page}{1}
\makeatletter
\renewcommand{\theequation}{A\arabic{equation}}
\renewcommand{\thefigure}{A\arabic{figure}}
\renewcommand{\bibnumfmt}[1]{[A#1]}
\renewcommand{\citenumfont}[1]{A#1}
%%%%%%%%%% Prefix a "S" to all equations, figures, tables and reset the counter %%%%%%%%%%

%%%%%%%%%%%%%%%%%%%%%%%%%%%%%%%%%%%%%%%%%%%%%%%%%%%%%%%%%%%%%%%%%%%%%%%%%%%
%%%%%%%%%%%%%%%%%%%    Proof of Entropy-Energy relation    %%%%%%%%%%%%%%%%
%%%%%%%%%%%%%%%%%%%%%%%%%%%%%%%%%%%%%%%%%%%%%%%%%%%%%%%%%%%%%%%%%%%%%%%%%%%
\section{Proof of Eqs.~(\ref{lndetrho}) and (\ref{E0S})}

We begin by proving Eq.~(\ref{lndetrho}). Using $D_n(T)=-4i\be \omega_n \check{\rho}(k)$ and Eq.~(\ref{E0det}), one obtains
\eq{
E_0 = - \lim_{T \to 0}\sum_{\bs{k}}\sum^{\infty}_{n=-\infty} T\ln \det \check{\rho}(k)+\ma{B}
}
where
\eq{\label{Bterm}
\ma{B} = - \lim_{T \to 0}\sum_{\bs{k}}\sum^{\infty}_{n=-\infty}T\ln\det(-4i\be(2n+1)\pi T I)\,.
}
where $I$ is the $4\times 4$ identity matrix. Simplifying $\ma{B}$, we get
\ea{
\ma{B} & = - \lim_{T \to 0}\sum_{\bs{k}}\sum^{\infty}_{n=-\infty}T\ln\det(-4i(2n+1)\pi I) \nonumber \\
& =  \lim_{T \to 0}\sum_{\bs{k}}\sum^{\infty}_{n=-\infty}-4T\ln(4(2n+1)\pi)\,.
}
We need to explicitly make the summation in $n$ before we take the limit $T \to 0$ since the total series may diverge. To make sense of the infinite summations, we will use zeta function regularization. Recall the Dirichlet series property
\eq{\label{Diri}
\sum^{\infty}_{n=1} (2n-1)^{-s} = (1-2^{-s})\zeta(s)
}
where we have the Riemann zeta function defined in the usual way as
\eq{
\zeta(s) = \sum^{\infty}_{n=1} n^{-s}\,.
}
Using the derivative
\eq{\label{zetads}
\frac{d}{ds}\left(\sum^{\infty}_{n=1} (2n-1)^{-s}\right) = - \sum^{\infty}_{n=1} (2n-1)^{-s} \ln(2n-1) 
}
together with Eq.~(\ref{Diri}), we obtain
\eq{\label{zetaOdd}
\sum^{\infty}_{n=1} (2n-1)^{-s} \ln(2n-1) = -2^{-s} \ln(2) \zeta(s) + (1-2^{-s}) \zeta'(s)\,.
}
Since
\eq{\label{zetapm}
\sum^{\infty}_{n=1} (2n-1)^{-s} = \sum^{\infty}_{n=1} (2n+1)^{-s} + 1
}
we also have that, from Eq.~(\ref{zetads}),
\eq{
\frac{d}{ds}\left(\sum^{\infty}_{n=1} (2n-1)^{-s}\right) = \frac{d}{ds}\left(\sum^{\infty}_{n=1} (2n+1)^{-s}\right)
}
and so
\eq{
\sum^{\infty}_{n=1} (2n+1)^{-s} \ln(2n+1) = \sum^{\infty}_{n=1} (2n-1)^{-s} \ln(2n-1) \,.
}
One final useful identity is
\eq{\label{npm}
\sum^{\infty}_{n=-\infty} (2n+1)^{-s} = (1-2^{-s})\zeta(s)(1+(-1)^{-s})\,,
}
which can be derived directly from Eqs.~(\ref{Diri}) and (\ref{zetapm}). In particular, for $s \in \mathbb{N}$, we have
\eq{\label{npm}
\sum^{\infty}_{n=-\infty} (2n+1)^{-s} =
\begin{cases}
  2(1-2^{-s})\zeta(s)\,, & \text{for $s$ even}\\
  0 & \text{for $s$ odd}\\
\end{cases}
}
where the identity for odd $s$ can be checked directly by explicit expansion of the series. We thus have
\ea{
\sum^{\infty}_{n=-\infty} \ln\left(4(2n+1)\pi\right) &  = \ln\left(4\pi\right) + 2\ln\left(4\pi\right)\sum^{\infty}_{n=1} 1  + \sum^{\infty}_{n=1} \ln\left(2n+1\right) + \sum^{\infty}_{n=1} \ln\left(-2n+1\right) \\
& = \ln\left(4\pi\right) + (2\ln\left(4\pi\right)+i\pi)\zeta(0) + 2\sum^{\infty}_{n=1} (2n-1)^{-s} \ln(2n-1) \bigg|_{s=0} \\
& = \ln(2)-\frac{i \pi}{2}
}
were we used $\zeta(0)=-1/2$. We can now safely take the limit $T \to 0$ in Eq.~(\ref{Bterm}), obtaining
\eq{\label{Aresult}
\ma{B} = -\lim_{T \to 0} \sum^{\infty}_{n=-\infty} 4T\ln\left(4(2n+1)\pi\right) = -\lim_{T \to 0}4T\left(\ln(2)-\frac{i \pi}{2}\right) = 0\,.
}
This concludes the proof of Eq.~(\ref{lndetrho}).

Moving on to the proof of Eq.~(\ref{E0S}), we start by proving the important intermediate relation
\eq{\label{trEq}
\ln\left(\textrm{Tr}[e^{-\be_D H}]\right) = S[\check{\rho}] + \be_D \om_{\bs{k}}\,.
}
Using Eq.~(\ref{rhoT}) and the fact that $\ln(AB)=\ln(A)+\ln(B)$ is $A$ and $B$ are commuting matrices, we obtain
\ea{
S[\check{\rho}] & = -\textrm{Tr}[\check{\rho} \ln(\check{\rho})] \nonumber \\
& = -\textrm{Tr}[\check{\rho} \ln(e^{-\be_D H})] + \textrm{Tr}[\check{\rho}\ln(\textrm{Tr}[e^{-\be_D H}])] \nonumber \\
& = \be_D \textrm{Tr}[\check{\rho}H] + \ln(\textrm{Tr}[e^{-\be_D H}])
}
where we used $\textrm{Tr}[\check{\rho}]=1$. All that remains is to find $\textrm{Tr}[\check{\rho}H]$, where we recall $H = -\left( m \g^0 - k_i \g^i \g^0 \right)/r_{{\bs{k}}}$. We start by noting that $[\check{\rho},H]=0$, so they share an eigenbasis. Expressing $H$ in its eigenbasis, one may check it will have the diagonal form $\tilde{H} = (k^0,k^0,-k^0,-k^0)$, from which one obtains
\eq{
\textrm{Tr}[\check{\rho}H] = \textrm{Tr}\left[\frac{e^{-\be_D \tilde{H} }}{\textrm{Tr}[e^{-\be_D \tilde{H}}]}\tilde{H} \right] = \frac{k^0 (e^{-k^0 \be_D}-e^{k^0 \be_D})}{e^{-k^0 \be_D}+e^{k^0 \be_D}}\,.
}
Finally, using Eqs.~(\ref{Tvirtual}) and (\ref{rk}) in the above result, one readily obtains
\eq{
\textrm{Tr}[\check{\rho}H] = -\om_{\bs{k}}\,,
}
from which Eq.~(\ref{trEq}) follows. Using now Eq.~(\ref{trEq}) and inserting Eq.~(\ref{rhoT}) in Eq.~(\ref{E0det}), one can write the vacuum energy $E_0$ as
\ea{\label{E0ABC}
E_0 & = - \lim_{T \to 0} \sum_{\bs{k}} T \sum^{\infty}_{n=-\infty} \left( 4\ln\left(4(2n+1)\pi\right) - 4S[\check{\rho}(k)] - 4\be_D \om_{\bs{k}} \right) \\
& = \sum_{\bs{k}} \left( A + B + C \right)
}
with
\ea{ 
A & = \lim_{T \to 0} \sum^{\infty}_{n=-\infty} -4T\ln\left(4(2n+1)\pi\right)\,, \label{AA} \\ 
B & = \lim_{T \to 0} \sum^{\infty}_{n=-\infty} 4T S[\check{\rho}(k_n)]\,, \label{BB} \\ 
C & = \lim_{T \to 0} \sum^{\infty}_{n=-\infty} 4T \be_D \om_{\bs{k}}\,. \label{CC}
}
Note that the factor of $e^{-\be_D H}$ in the numerator of Eq.~(\ref{rhoT}) does not contribute to the vacuum energy since ${\det(e^{-\be_D H})=1}$. We also have $A=\ma{B}$, so we know that $A=0$. Focusing on the quantity $C$, the summation of interest is
\eq{\label{Cpart}
\sum^{\infty}_{n=-\infty} \be_D \om_{\bs{k}} = \frac{1}{2}\sum^{\infty}_{n=-\infty} r_{{\bs{k}_n}} \ln\left(\frac{1+r_{{\bs{k}_n}}}{1-r_{{\bs{k}_n}}}\right) = \sum^{\infty}_{n=-\infty} \frac{1}{2i\al_n} \ln\left(\frac{i\al_n+1}{i\al_n-1}\right)
}
where we defined
\eq{\label{rkandalpha}
r_{{\bs{k}_n}} = \frac{1}{i\al_n}\,, \quad \al_n = \frac{(2n+1)\pi T}{\om_{\bs{k}}}\,.
}
Now, for $z\in \mathbb{C}$ we have \cite{Gradshteyn:2014}
\eq{
\frac{1}{2} \ln\left(\frac{z+1}{z-1}\right) = \textrm{arccoth}(z)
}
and for $z=i x$, $x\in \mathbb{R}$, we have
\eq{
\textrm{arccoth}(i x) = -i \textrm{arccot}(x)\,.
}
We can now use the series expansion around $x=0$ \cite{Gradshteyn:2014}
\eq{
\textrm{arccot}(x) = \frac{\pi}{2} - \sum^{\infty}_{p=0} \frac{(-1)^p x^{2p+1}}{2p+1}
}
to conclude that
\eq{
\frac{1}{2z} \ln\left(\frac{z+1}{z-1}\right) = -\frac{\pi}{2x} + \sum^{\infty}_{p=0} \frac{(-1)^p x^{2p}}{2p+1}\,.
}
Considering $x=\al_n$, applying the above expansion to Eq.~(\ref{Cpart}) and taking into account Eq.~(\ref{npm}), we obtain
\ea{
\sum^{\infty}_{n=-\infty} \be_D \om_{\bs{k}} & = -\frac{\omega_{\bs{k}}}{2T}\sum^{\infty}_{n=-\infty}(2n+1)^{-1} + \sum^{\infty}_{p=0} \frac{(-1)^p}{2p+1} \left(\frac{\pi T}{\om_{\bs{k}}}\right)^{2p}\sum^{\infty}_{n=-\infty}(2n+1)^{2p} \nonumber \\
& = \sum^{\infty}_{p=0} \frac{(-1)^p}{2p+1} \left(\frac{\pi T}{\om_{\bs{k}}}\right)^{2p} 2(1-2^{-2p})\zeta(-2p) \nonumber \\
& = 0 \label{Cis0}
}
where we used the fact that $\zeta(-2p)=0$ for non-zero positive integer $p$. We thus conclude that the term ${C=\lim_{T \to 0} \sum^{\infty}_{n=-\infty} 4T \be_D \om_{\bs{k}}}$ of the vacuum energy identically vanishes, even if we take finite temperatures.

Finally, we calculate the quantity $B$. To begin with, note that
\eq{
S[M] = - \textrm{Tr}[M \ln M] = - \sum_{j} \la_j \ln(\la_j)
}
where $\la_j$ are the eigenvalues of $M$. Since one can show that the eigenvalues of $\check{\rho}(k)$ are given by
\eq{
\frac{1\pm r_{\bs{k}}}{4}
}
each with multiplicity 2, this implies that
\eq{\label{explicitS}
S[\check{\rho}(k)] = -\left(\frac{1+r_{\bs{k}}}{2}\right) \ln\left(\frac{1+r_{\bs{k}}}{4}\right) - \left(\frac{1-r_{\bs{k}}}{2}\right) \ln\left(\frac{1-r_{\bs{k}}}{4}\right)\,.
}
Rearranging this, we obtain the relation
\ea{
B & = \lim_{T \to 0} \sum^{\infty}_{n=-\infty} 4T S[\check{\rho}(k_n)] \nonumber \\
& = -2\lim_{T \to 0} \sum^{\infty}_{n=-\infty} T \left\{\ln\left(\frac{1+r_{\bs{k}}}{4}\right) +  \ln\left(\frac{1-r_{\bs{k}}}{4}\right) +r_{\bs{k}}\ln\left(\frac{1+r_{\bs{k}}}{1-r_{\bs{k}}}\right)\right\} \nonumber \\
& = -2\lim_{T \to 0} \sum^{\infty}_{n=-\infty} T\left(\ln\left(\frac{1+r_{\bs{k}_n}}{4}\right) + \ln\left(\frac{1-r_{\bs{k}_n}}{4}\right)\right) - C\nonumber \\
& = -2\lim_{T \to 0} \sum^{\infty}_{n=-\infty} T\left(\ln\left(\frac{1+r_{\bs{k}_n}}{4}\right) + \ln\left(\frac{1-r_{\bs{k}_n}}{4}\right)\right) \label{Bfinal}
}
where we used Eqs.~(\ref{Tvirtual}) and (\ref{CC}). Using Eq.~(\ref{Aresult}) and Eq.~(\ref{Bfinal}), we obtain
\eq{\label{ABCfinal}
A + B + C = -2\lim_{T \to 0} \sum^{\infty}_{n=-\infty} T\left(\ln\left(\frac{1+r_{\bs{k}_n}}{4}\right) + \ln\left(\frac{1-r_{\bs{k}_n}}{4}\right)\right)\,.
}
In other words, we obtain
\eq{\label{finalE0}
E_0 = 4 \lim_{T \to 0} \sum_{\bs{k}}\sum^{\infty}_{n=-\infty} T S[\check{\rho}(k_n)]
}
where the sum assumes only a regularization. Note that the final expression for $E_0$ is divergence free in the $n$ summation, with no need for cancellation of infinite terms. Although we used zeta function techniques, the above result is valid for any choice of regularization. This concludes the proof.

As a sanity check of our result, one may use the identities (shown here without proof)
\ea{
\sum^{\infty}_{n=-\infty} (2n+1\pm a)^{-s} & = 2^{-s}\zeta\left(s,\frac{1\pm a}{2}\right) + (-2)^{-s}\zeta\left(s,-\frac{1\pm a}{2}\right) - \left(1\pm a\right)^{-s} \label{Hsum} \\
\zeta(0,a) & = \frac{1}{2}-a \label{H0} \\
\frac{\dd}{\dd s}\zeta(s,a) \bigg|_{s=0} & = \ln\Gamma(a) - \frac{1}{2}\ln(2\pi) \label{Hds}
}
to show that
\eq{\label{T0limit}
\lim_{T \to 0} \sum^{\infty}_{n=-\infty} T S[\check{\rho}(k_n)] = -\frac{1}{2}\lim_{T \to 0} \sum^{\infty}_{n=-\infty} T\left(\ln\left(\frac{1+r_{\bs{k}_n}}{4}\right) + \ln\left(\frac{1-r_{\bs{k}_n}}{4}\right)\right) = -\frac{\om_{\bs{k}}}{2}
}
and so, from Eq.~(\ref{E0ABC}) and Eq.~(\ref{ABCfinal}), we recover $E_0=-2\sum_{\bs{k}} \om_{\bs{k}}$, as expected. We can prove Eq.~(\ref{T0limit}) by using Eqs.~(\ref{explicitS}), (\ref{Cis0}) and (\ref{trEq}) and noting that
\ea{
-2\sum^{\infty}_{n=-\infty} S[\check{\rho}(k_n)] & = \sum^{\infty}_{n=-\infty} \left(\ln\left(\frac{1+r_{\bs{k}_n}}{4}\right) + \ln\left(\frac{1-r_{\bs{k}_n}}{4}\right)\right) \nonumber \\
& = \sum^{\infty}_{n=-\infty} \bigg(-4\ln(2)-2\ln(2n+1) + \ln\left(2n+1+\al\right) + \ln\left(2n+1-\al\right)\bigg) \label{LHS}
}
where we defined $\al = 1/(i \al_{0})$. Using $\sum^{\infty}_{n=-\infty} 1 = \zeta(0)=-1/2$ and $\sum^{\infty}_{n=-\infty} \ln(2n+1) = \ln(2)$, along with Eqs.~(\ref{Hsum})-(\ref{Hds}), the above relation simplifies to
\ea{
-2\sum^{\infty}_{n=-\infty} S[\check{\rho}(k_n)] & = \sum^{1}_{p=0}\sum^{\infty}_{n=-\infty} \ln\left(2n+1+(-1)^p\al\right) \nonumber \\
& = -\frac{\dd}{\dd s}\left[\sum^{1}_{p=0}\sum^{\infty}_{n=-\infty} \left(2n+1+(-1)^p\al\right)^{-s}\right]_{s=0} \nonumber \\
& = -\sum^{1}_{p=0}\frac{\dd}{\dd s}\left[2^{-s}\zeta\left(s,\frac{1 +(-1)^p \al}{2}\right) + (-2)^{-s}\zeta\left(s,-\frac{1+(-1)^p \al}{2}\right) - \left(1+(-1)^p \al\right)^{-s}\right]_{s=0} \nonumber \\
& = \sum^{1}_{p=0} \bigg\{\zeta\left(0,\frac{1 +(-1)^p \al}{2}\right)\ln(2) + \zeta\left(0,-\frac{1 +(-1)^p \al}{2}\right)(i \pi +\ln(2))-\ln(1+(-1)^p \al) \nonumber \\
& \hspace{20mm} - \frac{\dd}{\dd s}\zeta\left(0,\frac{1 +(-1)^p \al}{2}\right) - \frac{\dd}{\dd s}\zeta\left(0,-\frac{1 +(-1)^p \al}{2}\right)\bigg\} \nonumber \\
& = \sum^{1}_{p=0} \bigg\{i\pi+\ln(2)+\ln(2\pi)+(-1)^p\al\frac{i \pi}{2}-\ln(1+(-1)^p \al) -\ln\Gamma\left(\frac{1 +(-1)^p \al}{2}\right)-\ln\Gamma\left(-\frac{1 +(-1)^p \al}{2}\right)\bigg\} \nonumber \\
& = 2\pi i+2\ln(4)+2\ln(\pi)-\ln(1-\al^2) -\ln\bigg\{\Gamma\left(\frac{1}{2}+\frac{\al}{2}\right)\Gamma\left(\frac{1}{2}-\frac{\al}{2}\right)\Gamma\left(-\frac{1}{2}+\frac{\al}{2}\right)\Gamma\left(-\frac{1}{2}-\frac{\al}{2}\right)\bigg\} \nonumber
}

Now we may define $b=\omega_{\bs{k}}/(2\pi T)=\be \omega_{\bs{k}}/(2\pi)$, such that $\al = -2i b$ and use the relations
\ea{
\Gamma\left(\frac{1}{2}+i a\right)\Gamma\left(\frac{1}{2}-i a\right) & = \frac{\pi}{\cosh(\pi a)} \\
\Gamma\left(-\frac{1}{2}+i a\right)\Gamma\left(-\frac{1}{2}-i a\right) & = \frac{\pi}{\cosh(\pi a)}\left(\frac{1}{4}+a^2\right)^{-1}
}
to obtain
\ea{
-2\sum^{\infty}_{n=-\infty} S[\check{\rho}(k_n)] & = 2\pi i+2\ln(4)+2\ln(\pi)-\ln(1+4b^2) -\ln\bigg\{\Gamma\left(\frac{1}{2}+ib\right)\Gamma\left(\frac{1}{2}-ib\right)\Gamma\left(-\frac{1}{2}+ib\right)\Gamma\left(-\frac{1}{2}-ib\right)\bigg\} \nonumber \\
& = 2\pi i+2\ln(4)+2\ln(\pi)-\ln(1+4b^2) -\ln\bigg\{\left(\frac{\pi}{\cosh(\pi b)}\right)^2\left(\frac{1}{4}+b^2\right)^{-1}\bigg\} \nonumber \\
& = 2\pi i+\ln(4)+2\ln\left(\cosh\left(\frac{\be \omega_{\bs{k}}}{2}\right)\right) \nonumber \\
& = 2\pi i+\ln(4)+2\ln\left(\frac{e^{\be \omega_{\bs{k}}/2}+e^{-\be \omega_{\bs{k}}/2}}{2}\right) \nonumber \\
& = 2\pi i+\be \omega_{\bs{k}}+2\ln\left(1+e^{-\be \omega_{\bs{k}}}\right)\,. \label{RHS}
}
We thus obtain the summand of Eq.~(\ref{analyticZfermion}) up to a phase of $2\pi i$, which is of no physical consequence to the partition function $Z(T)$ in Eq.~(\ref{analyticZfermion}) since the term appears exponentiated. From Eq.~(\ref{LHS}) and Eq.~(\ref{RHS}), we may take the $T\to 0$ limit to prove Eq.~(\ref{T0limit}), as intended.

Finally, using Eqs.~(\ref{RHS}) and (\ref{analyticZfermion}) we find the result
\eq{\label{finiteTresult}
\sum_{\bs{k}}\sum^{\infty}_{n=-\infty} S[\check{\rho}(k_n)] = -\frac{\ln Z(T)}{4}\,,
}
which is valid for any temperature. The phase of $2\pi i$ in Eq.~(\ref{RHS}) obtained in the zeta function regularization contributes to the partition function $Z(T)$ through a multiplication by $e^{2\pi i}=1$, thus it has no physical consequence and can be safely removed from Eq.~(\ref{finiteTresult}).

\end{document}